\documentclass[twocolumn,pre,twoside,showpacs,preprintnumbers,floatfix]{revtex4}

\usepackage{amsmath}
\usepackage{amssymb}
\usepackage{graphicx}
\usepackage{dcolumn}
\usepackage{bm}

\def\m#1{\mathrm{#1}}
\def\Eq#1{(\ref{eq:#1})}

\def\dps{\displaystyle}
\def\Fig#1{\ref{fig:#1}}

\def\epsilon{\varepsilon}
\def\theta{\vartheta}
\def\rho{\varrho}
\def\Int#1#2#3{\int\limits_{#1}\!\mathrm{d}^{#2}{#3}\;}

\def\vec#1{\mathbf{#1}}

\begin{document}

%-------------------------------------------------------------------------------

\title{Phase Diagrams of Binary Mixtures of Oppositely Charged Colloids}

\author{Markus Bier}
\email{bier@mf.mpg.de}
\affiliation
{
   Max-Planck-Institut f\"{u}r Metallforschung, 
   Heisenbergstra{\ss}e 3,
   70569 Stuttgart,
   Germany, 
   and
   Institut f\"{u}r Theoretische und Angewandte Physik,
   Universit\"{a}t Stuttgart,
   Pfaffenwaldring 57,
   70569 Stuttgart,
   Germany
}

\author{Ren\'e van Roij}
\affiliation
{
   Institute for Theoretical Physics, 
   Utrecht University, 
   Leuvenlaan 4, 
   3584\,CE Utrecht, 
   The Netherlands
}

\author{Marjolein Dijkstra}
\email{m.dijkstra1@uu.nl}
\affiliation
{
   Soft Condensed Matter,
   Debye Institute for NanoMaterials Science
   Utrecht University,
   Princetonplein 5,
   3584\,CC Utrecht,
   The Netherlands
}

\date{21 May, 2010}

\begin{abstract}
   Phase diagrams of binary mixtures of oppositely charged colloids are calculated theoretically. 
   The proposed mean-field-like formalism interpolates between the limits of a hard-sphere 
   system at high temperatures and the colloidal crystals which minimize Madelung-like energy 
   sums at low temperatures.
   Comparison with computer simulations of an equimolar mixture of oppositely charged,
   equally sized spheres indicate semi-quantitative accuracy of the proposed formalism.
   We calculate global phase diagrams of binary mixtures of equally sized spheres with opposite 
   charges and equal charge magnitude in terms of temperature, pressure, and composition.
   The influence of the screening of the Coulomb interaction upon the topology of the phase 
   diagram is discussed.
   Insight into the topology of the global phase diagram as a function of the system parameters
   leads to predictions on the preparation conditions for specific binary colloidal crystals. 
\end{abstract}

\pacs{64.70.pv, 64.70.D-, 64.70.K-}

\maketitle

%-------------------------------------------------------------------------------

\section{Introduction}

Colloidal suspensions are interesting experimental realizations of many-particle 
systems because, in contrast to molecules, they are directly accessible to observations and
the interactions can be tuned in a wide range by means of the preparation procedure and the 
experimental conditions \cite{Habdas2002,Tata2006,Prasad2007,Liang2007}.
The flexibility of colloidal suspensions, however, implies many system parameters to be
specified, such as size and shape of the colloids as well as strength and functional form of 
the interaction potential \cite{Dijkstra2001,Yethiraj2003,Leunissen2005,Shevchenko2006}. 
In addition, the number of parameters increases enormously if one considers mixtures of different
colloidal species instead of a pure system of only one type of colloid.
Moreover, the dimension of the phase diagram increases in parallel with the number of colloidal
species in the suspension.
The high dimensionality of the parameter space and the phase diagrams call for a simple theory
to get an overview of the global phase behavior of colloidal suspensions because a systematic
scan by means of experiments or computer simulations is precluded.

In the present work such a simple theory is formulated for the case of binary mixtures of 
spherical colloids of the same radius and opposite charges of equal magnitude.
The global phase diagrams for this setting are in terms of the temperature, the pressure, and
the composition and there is only one additional parameter describing the screening of the Coulomb
interaction.
The reason to consider this comparatively low-dimensional problem here is that the reliability of
the approach should be assessed with respect to available results on special cases obtained by 
means of computer simulations \cite{Hynninen2003,Hynninen2006a} and Madelung sum calculations 
\cite{Leunissen2005,Maskaly2006,vdBerg2009}. 
An extension of the method to multi-component, size- and charge-polydisperse colloidal 
suspensions is readily achieved along the same lines.

The proposed mean-field-like formalism correctly interpolates between the hard-sphere limit at
high temperatures and a Madelung-type description at low temperatures. 
As a mean-field theory, the present approach underestimates fluctuations such that critical 
phenomena and quantitative aspects of phase transitions are covered in lowest order only.
The over-all topology of the phase diagram, in particular the types of phases present, however, 
is not expected to be influenced qualitatively.
The simple and transparent formalism turns out to be in semi-quantitative agreement with
the computer simulation results of Ref.~\cite{Hynninen2006a}; this gives confidence that a 
possible extension to multi-component, size- and charge-polydisperse colloidal suspensions leads
to a reliable description. 
 
The model, the formalism, and further technical details are introduced in
Sec.~\ref{sec:formalism}.
Results are discussed in Sec.~\ref{sec:results} and Sec.~\ref{sec:summary} concludes with a 
summary.

\vfill

%-------------------------------------------------------------------------------

\section{\label{sec:formalism}Formalism}

\subsection{Model}

Consider a binary mixture of $N_1$ positively charged colloids and $N_2$ negatively charged 
colloids in a solvent of volume $V$, temperature $T$, Debye screening length $\kappa^{-1}$, and dielectric
constant $\epsilon$.
For simplicity we restrict attention to equally sized spherical colloids of radius $a$ with exactly
opposite valency $Z_1=Z>0$ and $Z_2=-Z<0$ for the positively and negatively charged colloids, 
respectively.
For later use we define the sign of the colloidal charge $q_i = Z_i/Z, i\in\{1,2\}$.
The thermodynamic state can be characterized by the total volume fraction $\phi= 4\pi a^3 N/(3V)$
and the mole fraction of the positively charged colloids $x_1 = x = N_1/N$ where $N=N_1 + N_2$ 
is the total number of colloids; the mole fraction of the negatively charged colloids is given 
by $x_2 = 1-x$.
The osmotic pressure $p$ will be reported in terms of the dimensionless combination 
$p^* = 4\pi pa^3/(3k_BT)$ where $k_B$ is the Boltzmann constant.
The effective colloidal interactions are assumed to be pairwise additive with hard-core and 
screened-Coulomb contributions but without Van der Waals attractions; these conditions are 
realized in index-matched solvents.
In terms of the dimensionless temperature $T^* = 4\pi\epsilon_0\epsilon k_BTa(1+\kappa a)^2/
(Ze)^2$ with the permittivity of the vacuum $\epsilon_0$ and the elementary charge $e$, the pair
potential can thus be written as
\begin{equation}
   \beta U_{ij}(r) =
   \left\{\begin{array}{ll}
      \infty                                                   & , r < 2a \\
      \dps\frac{q_iq_j}{T^*}\frac{\exp(-\kappa a(r/a-2))}{r/a} & , r \geq 2a
   \end{array}\right.,
   \label{eq:U}
\end{equation}
where $\beta = 1/(k_BT)$ is the inverse temperature.

\subsection{Gibbs free energy}

The phase behavior of the system under consideration will be calculated from the Gibbs free
energy per particle per $k_BT$
\begin{equation}
   g(T^*, p^*, x) = f(T^*,\Phi(T^*,p^*,x),x) + \frac{p^*}{\Phi(T^*,p^*,x)},
   \label{eq:g}
\end{equation}
where $f(T^*,\phi,x)$ is the Helmholtz free energy per particle per $k_BT$ and $\Phi(T^*,p^*,x)$ is the
total volume fraction for given $(T^*,p^*,x)$, which is implicitly defined by means of the 
equation of state
\begin{equation}
   p^* = \phi^2\frac{\partial f}{\partial \phi}\bigg|_{\phi=\Phi(T^*,p^*,x)}.
   \label{eq:eos}
\end{equation}
Below $f(T^*,\phi,x)$, and hence $g(T^*,p^*,x)$ using Eqs.~\Eq{g} and \Eq{eos}, is calculated for (i) fluid,
(ii) crystalline, and (iii) substitutionally disordered solid phases.
The expressions involve a priori unknown constants which are fixed by fitting to the well-known
hard-sphere fluid-crystal transition.

\subsubsection{Fluid phase}

The Helmholtz free energy per particle of the fluid phase is approximated by the analytical
solution of the mean spherical approximation (MSA) of hard-sphere Yukawa mixtures 
\cite{Ginoza1990}
\begin{eqnarray}
   f(T^*,\phi,x) 
   & = & 
   c_f+\sum_ix_i\ln(x_i)+\ln(\phi)+\frac{\phi(4-3\phi)}{(1-\phi)^2} \nonumber\\
   & &
   +\frac{B(\Gamma)}{2T^*} + \frac{\Gamma^2(\Gamma+3\kappa a)}{18\phi}
   \label{eq:ffluid}
\end{eqnarray}
with the integration constant $c_f$, which will be fixed below, the excess internal energy due to
the screened-Coulomb interaction $B(\Gamma)$ given by Eq.~(14) of Ref.~\cite{Ginoza1990}, and the 
effective screening strength $\Gamma$ being the non-negative solution of the 6th-order algebraic 
equation~(10) of Ref.~\cite{Ginoza1990}, which has to be solved numerically.
Upon $T^*\rightarrow\infty$ one finds $\Gamma\rightarrow0$ such that
Eq.~\Eq{ffluid} leads to the Carnahan-Starling equation of state \cite{Carnahan1969}.

\subsubsection{Crystalline phases}

The Helmholtz free energy per particle of crystalline phases is approximated by the following 
mean-field functional of the one-particle density profiles $\rho_i, i\in\{1,2\}$, where 
$\rho_i(\vec{r})$ is the number density of positively ($i=1$) or negatively ($i=2$) charged 
colloids at position $\vec{r}$ in space: 
\begin{eqnarray}
   & &
   f(T^*,\phi) - f^\m{hs}(\phi)
   \label{eq:fsolid}\\
   & = &
   \frac{1}{2N}\Int{V}{3}{r}\Int{V(\vec{r})}{3}{r'}\sum_{i,j}
   \rho_i(\vec{r})\rho_j(\vec{r'})\beta U_{ij}(|\vec{r'}-\vec{r}|).
   \nonumber
\end{eqnarray}
Here $f^\m{hs}$ is the hard-sphere Helmholtz free energy per particle per $k_BT$ in the crystalline
phase, $V$ is the system volume, and $V(\vec{r}) = \{\vec{r'}\in V: |\vec{r'}-\vec{r}| \geq 2a\}$.
Note that the concentration $x$ of a crystal is fixed by the crystal structure and not a free 
parameter.
Given the crystal structure $\mathcal{C} = \mathcal{C}_1 \cup \mathcal{C}_2$ composed of the
sublattices $\mathcal{C}_i, i\in\{1,2\}$ of positively ($i=1$) and negatively ($i=2$) charged 
colloids, an approximation to the density profiles is 
\begin{equation}
   \rho_i(\vec{r}) = \sum_{\vec{s}\in\mathcal{C}_i}\delta(\vec{r}-\vec{s}),
   \label{eq:densitycrystal}
\end{equation}
where $\delta$ is the Dirac delta.
The energetic contribution to the Helmholtz free energy per particle, given by the term on the 
right-hand side of Eq.~\Eq{fsolid}, follows directly from Eqs.~\Eq{U} and \Eq{densitycrystal};
it is the analog for the screened Coulomb potential Eq.~\Eq{U} of the well-known Madelung energy
\cite{Kittel1996}. 
The hard-sphere Helmholtz free energy per particle, $f^\m{hs}$, is approximated by integrating
the free-volume equation of state corresponding to crystal structure $\mathcal{C}$ of 
closed-packed packing fraction $\phi_\m{cp}$ \cite{Wood1952} with respect to the packing 
fraction $\phi$.

Finally the Helmholtz free energy per particle of crystal structure $\mathcal{C}$ reads
\begin{eqnarray}
   & &
   f(T^*,\phi) 
   \label{eq:fcrystal}\\
   & = &
   c_s + \ln(\phi) - 3\ln\Bigg(1-\left(\frac{\phi}{\phi_\m{cp}}\right)^{1/3}\Bigg) 
   \nonumber\\
   & & 
   +\frac{1}{2K_cT^*}\sum_{i,j}q_iq_j\sum_{(\vec{s},\vec{s'})\in\mathcal{C}^*_{ij}} 
   \frac{\exp(-\kappa a(|\vec{s}-\vec{s'}|/a - 2))}{|\vec{s}-\vec{s'}|/a}
   \nonumber
\end{eqnarray}
with the number of colloids per unit cell $K_c$ and the abbreviation 
$\mathcal{C}^*_{ij} = \{(\vec{s},\vec{s'})\in\mathcal{C}_i\times\mathcal{C}_j:
\vec{s'}\in W_\mathcal{C}, \vec{s} \not= \vec{s'}\}$, where $W_\mathcal{C}$ denotes the unit cell.
The integration constant $c_s$ in Eq.~\Eq{fcrystal} is \emph{independent} of the 
closed-packed packing fraction $\phi_\m{cp}$ and the crystal structure $\mathcal{C}$.

\subsubsection{Substitutionally disordered solid phase}

The density profiles of a crystal structure $\mathcal{C}$ with the sites occupied randomly by 
positively and negatively charged colloids with probability $x_1$ and $x_2$, respectively, are 
approximated, in place of Eq.~\Eq{densitycrystal}, by
\begin{equation}
   \rho_i(\vec{r}) = x_i\sum_{\vec{s}\in\mathcal{C}}\delta(\vec{r}-\vec{s}).
   \label{eq:densityamorph}
\end{equation}
Therefore, the Helmholtz free energy per particle is approximated by
\begin{eqnarray}
   & &
   f(T^*,\phi,x)  
   \label{eq:famorph}\\
   & = & 
   c_s + \sum_i x_i\ln(x_i) + \ln(\phi) 
   - 3\ln\Bigg(1-\left(\frac{\phi}{\phi_\m{cp}}\right)^{1/3}\Bigg)
   \nonumber\\
   & & 
   +\frac{(2x-1)^2}{2K_cT^*}\sum_{(\vec{s},\vec{s'})\in\mathcal{C}^*}
   \frac{\exp(-\kappa a(|\vec{s}-\vec{s'}|/a - 2))}{|\vec{s}-\vec{s'}|/a}
   \nonumber
\end{eqnarray}
with $\mathcal{C}^*_{\phantom{ij}} = \{(\vec{s},\vec{s'})\in\mathcal{C}\times\mathcal{C}: 
\vec{s'}\in W_\mathcal{C}, \vec{s} \not= \vec{s'}\}$.
Note the ``entropy of mixing'' contribution due to the random occupation of sites and that
the screened-Coulomb energy vanishes at $x=1/2$ for the present choice of equally sized, oppositely
charged particles.

\subsubsection{\label{sec:freezing}Hard-sphere freezing}

In order to fix the values of the integration constants $c_f$ (see Eq.~\Eq{ffluid}) and 
$c_s$ (see Eqs.~\Eq{fcrystal} and \Eq{famorph}) it is required that hard-sphere freezing 
from a fluid to a random face-centered cubic (rfcc) phase, i.e., a substitutionally disordered 
solid, takes place in the limit $T^*\rightarrow\infty$ at 
$p^*_\m{coex} = 11.54(4)\pi/6 \approx 6.042$ (see Ref.~\cite{Noya2008}).
As the Gibbs free energies per particle of the fluid and the rfcc solid must be equal at 
coexistence, the two constants $c_f$ and $c_s$ are related by $c_f-c_s \approx 1.084$. 
As only differences of Gibbs free energies are relevant in order to determine the equilibrium 
state, we set $c_s=0$ without restriction of generality. 
The binodals of the fluid-rfcc solid coexistence region at high temperatures are found to be 
located at packing fractions $\phi_\m{f} \approx 0.4916, \phi_\m{rfcc} \approx 0.5547$, i.e., the 
coexistence region is in good agreement with that found in computer simulations \cite{Noya2008}, 
albeit slightly wider.

\subsection{Phase diagrams}

In order to calculate phase diagrams in terms of $T^*$, $p^*$, and $x$ a set of candidate solid
structures is chosen in the next section.
Let $\tilde{g}(T^*, p^*, x)$ be defined as the minimum of the Gibbs free energies per particle 
constructed in the previous subsection of the fluid, the candidate crystals, and the candidate 
substitutionally disordered solids for given $(T^*, p^*, x)$.
The equilibrium Gibbs free energy is the convex hull of $\tilde{g}(T^*, p^*, x)$ as a function 
of $x\in[0,1]$, which can be readily constructed numerically on a $x$-grid. 
During this calculation the coexistence regions of the phase diagrams can be identified as the 
set of state points where Maxwell's common tangent construction applies.

The Gibbs free energies per particle as defined in the previous subsection exhibit an unphysical
feature at low pressures, where for all temperatures the fluid becomes apparently unstable with
respect to an rfcc crystal; the reason for this phenomenon is related to the well-known incorrect 
low-density asymptotics of the free-volume equation of state \cite{Wood1952}. 
In order to resolve this problem it is assumed that once a stable fluid state for given 
$(T^*,p^*,x)$ is found, a stable fluid state is also assumed for $(T^*,x)$ and all pressures smaller
than $p^*$.

%-------------------------------------------------------------------------------

\begin{table}[!t]
   \begin{tabular}{|c|c|c|c|}
      \hline
      structure          & c/d & $K_c$ & $\phi_\m{cp}$                           \\
      \hline
      $\m{CaF_2}$        & c   & $3$   & $3\pi\sqrt{3}/32$                       \\
      $\m{CsCl}$         & c   & $2$   & $\pi\sqrt{3}/8$                         \\
      $\m{CuAu}$         & c   & $2$   & $\pi(2+(c/a)^2)\sqrt{1+2(a/c)^2}/24$ \\
      $\m{Cu_3Au}$       & c   & $4$   & $\pi\sqrt{2}/6$                         \\
      $\m{LS_6^\m{fcc}}$ & c   & $7$   & $7\pi\sqrt{2}/48$                       \\
      $\m{NaCl}$         & c   & $2$   & $\pi/6$                                 \\
      $\m{NbP}$          & c   & $8$   & $\pi\sqrt{2}/6$                         \\
      $\m{rbcc}$         & d   & $1$   & $\pi\sqrt{3}/8$                         \\
      $\m{rfcc}$         & d   & $1$   & $\pi\sqrt{2}/6$                         \\
      \hline
   \end{tabular}
   \caption{\label{tab:candidates}Candidate solid structures chosen on the basis of 
            Refs.~\cite{Leunissen2005,Hynninen2003,Hynninen2006a,Maskaly2006,vdBerg2009}.
            For detailed structure information see Ref.~\cite{CrystStruct}.
            The second column indicates a crystal (``c'') or a substitutionally disordered
            solid (``d'').
            The third and the fourth columns give the numbers of particles per unit cell 
            $K_c$ and the closed-packed packing fractions $\phi_\m{cp}$, respectively.
            The $\m{CuAu}$ structure with the aspect ratio $c/a\in[1,\sqrt{2}]$ of the 
            underlying tetragonal lattice degenerates to the $\m{CsCl}$ structure in the 
            case $c/a=1$.
            The $\m{LS_6^\m{fcc}}$ structure has been introduced in Ref.~\cite{Hynninen2006b}.
            The $\m{NbP}$ structure was called ``tetragonal'' in Ref.~\cite{Hynninen2006a}.}
\end{table}

\begin{figure*}[!t]
   \includegraphics{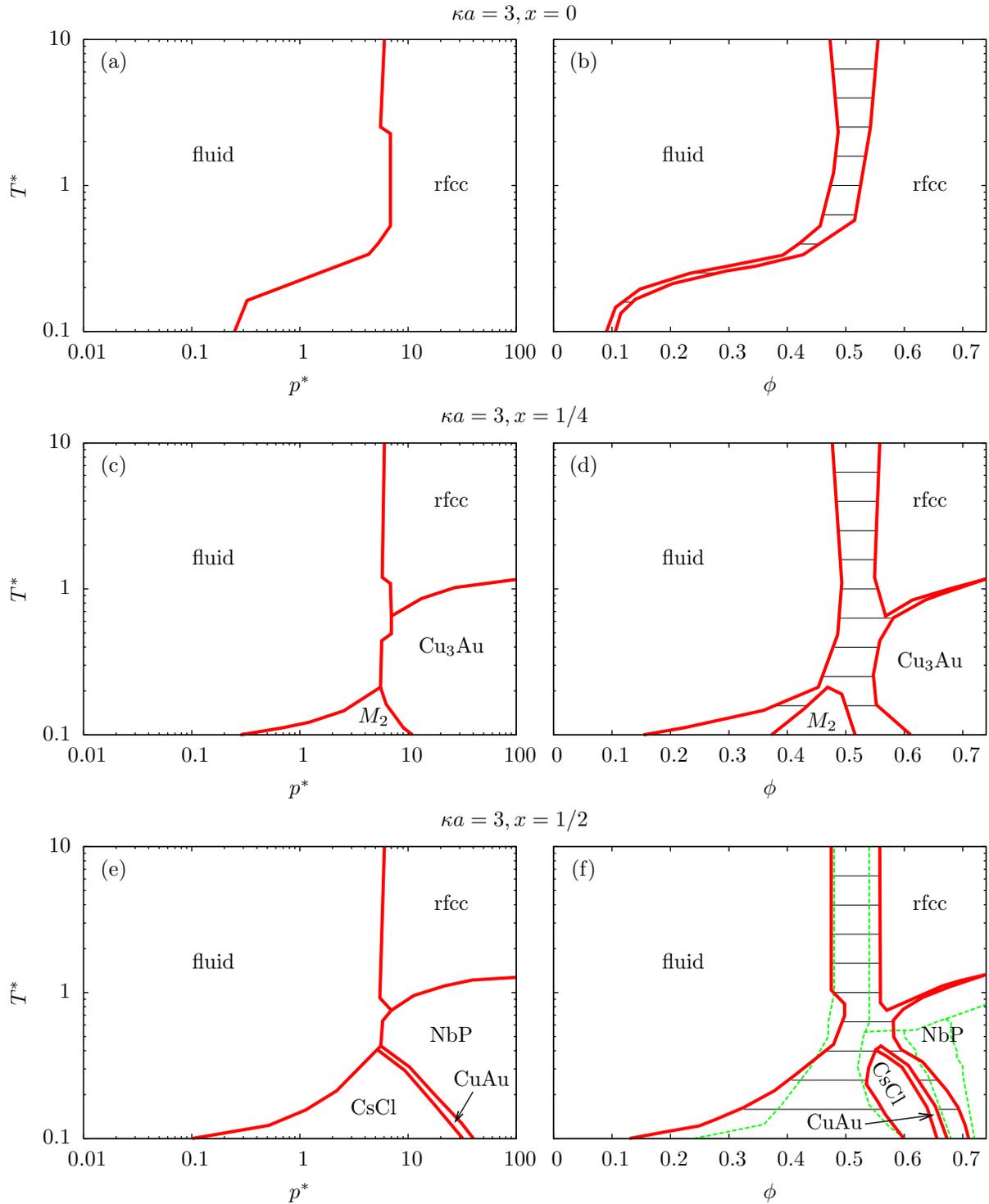}
   \caption{\label{fig:Tpphi3}Phase diagrams for $\kappa a = 3$ at compositions $x$ in terms 
            of $(T^*, p^*)$ and $(T^*, \phi)$. 
            The green dashed lines in panel (f) reproduce the phase diagram of the computer 
            simulation study of Ref.~\cite{Hynninen2006a}.
            Labels of pure solid phases correspond to Tab.~\ref{tab:candidates}. 
            Panels~(c) and (d) exhibit a two-phase coexistence region $M_2=\m{CsCl}+\m{rfcc}$.
            Horizontal thin lines in panels (b), (d), and (f) are tie lines.}
\end{figure*}

\begin{figure*}[!t]
   \includegraphics{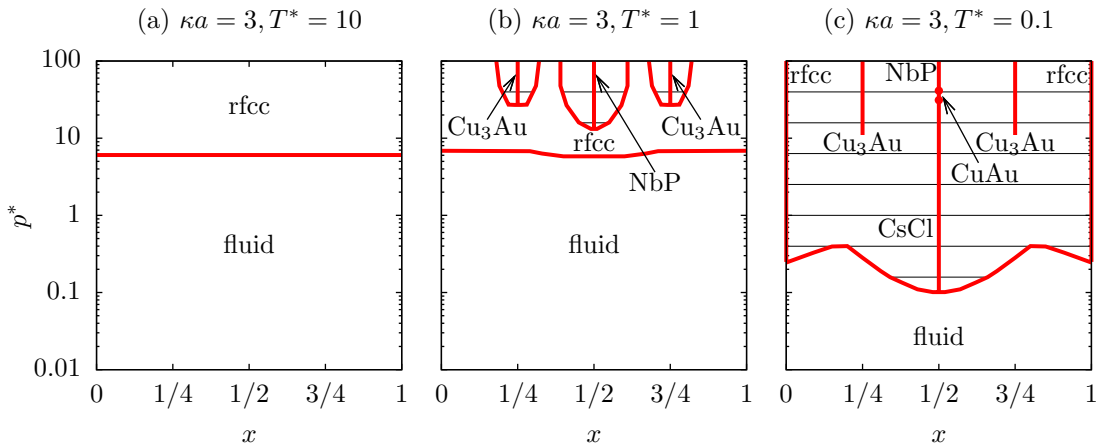}
   \caption{\label{fig:px3}Phase diagrams for $\kappa a = 3$ in terms of $(p^*,x)$ at 
            temperatures (a) $T^*=10$, (b) $T^*=1$, and (c) $T^*=0.1$ (see also 
            Fig.~\Fig{Tpphi3}).
            Horizontal thin lines in panels (b) and (c) are tie lines.}
\end{figure*}

\begin{figure*}[!t]
   \includegraphics{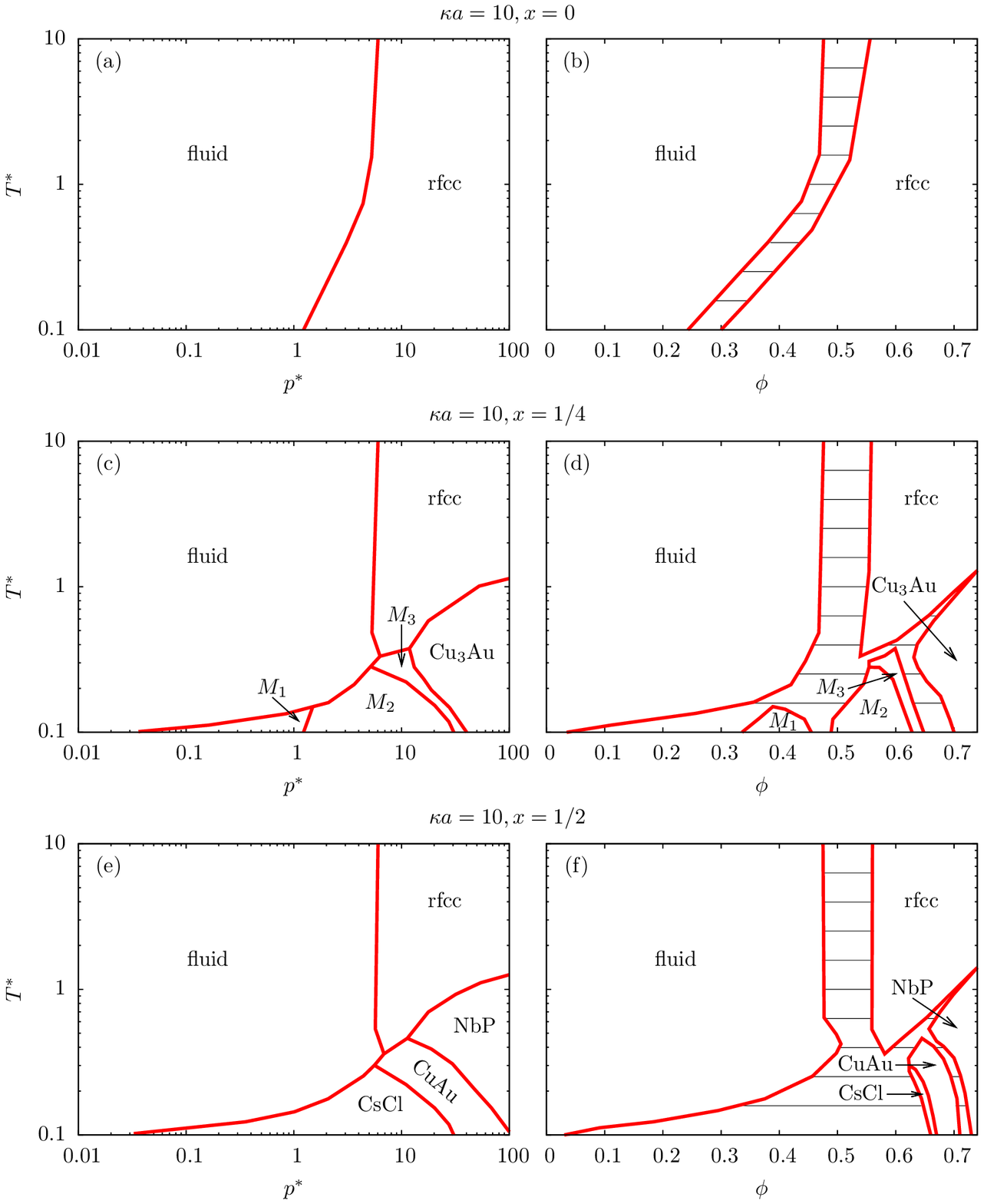}
   \caption{\label{fig:Tpphi10}Phase diagrams for $\kappa a = 10$ at compositions $x$ in terms
            of $(T^*, p^*)$ and $(T^*, \phi)$.
            Labels of pure solid phases correspond to Tab.~\ref{tab:candidates}. 
            Panels~(c) and (d) exhibit two-phase coexistence regions $M_1=\m{CsCl}+\m{fluid}$,
            $M_2=\m{CsCl}+\m{rfcc}$, and $M_3=\m{CuAu}+\m{rfcc}$.
            Horizontal thin lines in panels (b), (d), and (f) are tie lines.}
\end{figure*}

\section{\label{sec:results}Results and Discussion}

Table~\ref{tab:candidates} lists the candidate solid structures considered here.
This choice is based on the structures found in computer simulation studies of the cases
$x=0$ (see Ref.~\cite{Hynninen2003}) and $\kappa a=3,x=1/2$ (see Ref.~\cite{Hynninen2006a}), 
as well as on Refs.~\cite{Leunissen2005,Maskaly2006,vdBerg2009}, where the limit $T^* \rightarrow 0$ 
is addressed by means of Madelung energy sums.
Moreover, the $\m{CsCl}$ (cesium chloride), $\m{CuAu}$ (copper gold), $\m{NaCl}$ (sodium 
chloride), and $\m{Cu_3Au}$ structures have been identified in experiments 
\cite{Leunissen2005,Shevchenko2006}.
Table~\ref{tab:candidates} indicates whether the solid is crystalline (``c'') or
substitutionally disordered (``d'') and it exhibits the numbers of particles per unit cell 
$K_c$ as well as the closed-packed packing fractions $\phi_\m{cp}$.
Detailed structure information can be found in Ref.~\cite{CrystStruct}.
The $\m{CuAu}$ structure, which is described by a tetragonal lattice of aspect ratio 
$c/a\in[1,\sqrt{2}]$ and a two-particle basis, degenerates to the $\m{CsCl}$ structure in the
case $c/a=1$.
The structure denoted by $\m{LS_6^\m{fcc}}$ was introduced in Ref.~\cite{Hynninen2006b} and the 
$\m{NbP}$ (niobium phosphide) structure was called ``tetragonal'' in Ref.~\cite{Hynninen2006a}.

\subsection{Case $\kappa a=3$}

The phase diagrams for the case $\kappa a = 3$ in terms of $(T^*,p^*)$ and $(T^*,\phi)$ at 
compositions $x=0$, $x=1/4$, and $x=1/2$ are displayed in Fig.~\Fig{Tpphi3}.
Thick solid lines represent phase boundaries whereas thin horizontal lines in $(T^*,\phi)$ 
diagrams (panels (b), (d), and (f)) are tie lines connecting coexisting states.
At low temperatures two-phase coexistence $M_2:=\m{CsCl}+\m{rfcc}$ is found (see Figs.~\Fig{Tpphi3}(c)
and (d)).

By construction a $\m{fluid}$-$\m{rfcc}$ transition takes place for any $\kappa a$ and $x$ in the limit 
$T^*\rightarrow\infty$ at the coexistence pressure $p^*_\m{coex}$ and volume fractions $\phi_\m{f}$ and
$\phi_\m{rfcc}$ (see Sec.~\ref{sec:freezing}).
For $T^*\rightarrow 0$ and $x=1/2$ a $\m{CsCl}$ crystal coexists with a dilute gas, in agreement with 
Ref.~\cite{vdBerg2009} because the free energy of the present formalism reduces to Madelung-like
energy sums in this limit.
A cut of the phase diagrams at composition $x=0$ is shown in Figs.~\Fig{Tpphi3}(a) and (b),
which involves merely the $\m{fluid}$ and the $\m{rfcc}$ structures, which is consistent with computer
simulation results \cite{Hynninen2003}.
Figures~\Fig{Tpphi3}(c) and (d) display a cut at $x=1/4$, with an additional $\m{Cu_3Au}$ 
phase as well as the two-phase coexistence region $M_2$.
The structures at composition $x=1/4$ differ from those at composition $x=3/4$ only by an 
exchange of the colloid species ($1\leftrightarrow2$).

In order to assess the reliability of the approach described in Sec.~\ref{sec:formalism}, 
Fig.~\Fig{Tpphi3}(f) compares the calculated phase diagram (red solid lines) in terms of 
$(T^*, \phi)$ for $\kappa a=3, x=1/2$ with that obtained by means of free energy calculations
using computer simulations in Ref.~\cite{Hynninen2006a} (green dashed lines).
Both studies agree in the predicted stable structures $\m{fluid}$, $\m{rfcc}$, $\m{CsCl}$, 
$\m{CuAu}$, and $\m{NbP}$.
The overall topology is similar for both approaches; however, the present formalism 
overestimates the stability of the $\m{NbP}$ structure leading to a $\m{fluid}$-$\m{NbP}$ 
transition (see Figs.~\Fig{Tpphi3}(e) and (f)) which is not observed in the computer simulation.
Moreover, our formalism leads to a $\m{fluid}$-$\m{CuAu}$ transition in a very narrow window around
$T^*\approx 0.4$ but no $\m{rfcc}$-$\m{CuAu}$ transition is found, whereas it is the opposite situation 
with the computer simulation results.
Agreement between mean-field theory and computer simulation is observed with respect to the order
of the phase transitions:
The $\m{CsCl}$-$\m{CuAu}$ transition is of second order because the $\m{CuAu}$ structure 
transforms continuously into $\m{CsCl}$ upon $c/a\rightarrow1$, and the other phase transitions
are of first order.
Both the $\m{rfcc}$-$\m{NbP}$ and the $\m{CuAu}$-$\m{NbP}$ phase transitions are described 
as ``weakly first-order'' in Ref.~\cite{Hynninen2006a}, whereas within our approach, only the 
$\m{rfcc}$-$\m{NbP}$ transition exhibits a very narrow but non-vanishing $\phi$-gap and the 
$\m{CuAu}$-$\m{NbP}$ transitions is strongly first-order (see Fig.~\Fig{Tpphi3}(f)).
The quantitative disagreement in the strength of the first-order $\m{CuAu}$-$\m{NbP}$ transition
can be understood on the basis of a smearing out of structural differences due to fluctuations, which are 
present in computer simulations but which are not fully accounted for by mean-field theories. 
This comparison between the formalism of Sec.~\ref{sec:formalism} and the computer simulation
study of Ref.~\cite{Hynninen2006a} for the special case $\kappa a=3, x=1/2$ shows that, apart 
from well-known defects of mean-field theories, the present approach is semi-quantitatively 
reliable.
Moreover, the simplicity of the present formalism gives computational advantages over computer 
simulations such that now complete phase diagrams in terms of $(T^*, p^*, x)$ as a function of
the parameter $\kappa a$ can be determined readily.

Figure~\Fig{px3} displays the phase diagram for $\kappa a=3$ in terms of $(p^*,x)$ for 
temperatures (a) $T^*=10$, (b) $T^*=1$, and (c) $T^*=0.1$.
At high temperatures (see Fig.~\Fig{px3}(a)) only $\m{fluid}$ and $\m{rfcc}$ structures are 
present and the $\m{fluid}$-$\m{rfcc}$ transition line becomes independent of the composition 
$x$ in the limit $T^*\rightarrow\infty$.
At lower temperatures (see Figs.~\Fig{px3}(b) and (c)) $\m{CsCl}$, $\m{CuAu}$, $\m{Cu_3Au}$, 
and $\m{NbP}$ crystal structures occur at fixed compositions. 
At low temperatures and pressures as well as strongly asymmetric mixtures (see 
Fig.~\Fig{px3}(c) for $p^*<1$ and $x<1/4$ or $x>3/4$) the MSA approximation applied to model
the fluid phase leads to an unphysical artifact which exhibits the apparent coexistence of
an almost pure $\m{rfcc}$ crystal with a less pure fluid.
The reason for this unphysical phenomenon is that under these conditions the MSA pair 
distribution function becomes negative such that an increasing repulsive interaction potential
leads to a more and more negative, i.e., attractive, contribution to the free energy.
However, outside of this range of the phase diagram MSA leads to a physically reasonable
description of the fluid phase.
The full phase diagram for $\kappa a=3$ in terms of $(T^*,p^*,x)$ can be inferred from the 
two-dimensional cuts in Figs.~\Fig{Tpphi3} and \Fig{px3}.

\subsection{Case $\kappa a=10$}

In order to study the changes of the phase diagram upon changing the screening strength 
$\kappa a$, the phase diagrams for various screening strengths $\kappa a\in\{2,3,5,10\}$ have 
been calculated.
The trends in the variations of the phase diagrams upon changing $\kappa a$ are found to be
monotonic such that it is sufficient in the following to consider the case $\kappa a = 10$.

Figures~\Fig{Tpphi10} displays the phase diagram for $\kappa a=10$ in terms of $(T^*,p^*)$ and 
$(T^*,\phi)$ at compositions $x=0$, $x=1/4$, and $x=1/2$.
Two-phase coexistence regions $M_1:=\m{CsCl}+\m{fluid}$, $M_2=\m{CsCl}+\m{rfcc}$, and 
$M_3:=\m{CuAu}+\m{rfcc}$ are present in Figs.~\Fig{Tpphi10}(c) and (d).
By comparison of Figs.~\Fig{Tpphi3} and \Fig{Tpphi10} one infers a shift of the 
$\m{rfcc}$ phase to lower temperatures upon increasing $\kappa a$.
This observation can be understood by the fact that the interaction potential is approaching 
the hard-sphere potential in the limit $\kappa a\rightarrow\infty$.
The $\m{Cu_3Au}$ phase in Figs.~\Fig{Tpphi3}(c) and \Fig{Tpphi10}(c) or Figs.~\Fig{Tpphi3}(d) 
and \Fig{Tpphi10}(d) shrinks upon increasing $\kappa a$, which is partly due to the growing 
$\m{rfcc}$ phase.
Figures~\Fig{Tpphi3}(e) and \Fig{Tpphi10}(e) or Figs.~\Fig{Tpphi3}(f) and \Fig{Tpphi10}(f)
exhibit an increasing temperature range of stability of the $\m{CuAu}$ structure upon 
increasing $\kappa a$.
As a consequence the $\m{NbP}$ phase, located in between the extending $\m{rfcc}$ and 
$\m{CuAu}$ phases, shrinks upon increasing $\kappa a$.
Moreover, upon increasing $\kappa a$, the $\m{fluid}$-$\m{NbP}$ transition disappears and an 
$\m{rfcc}$-$\m{CuAu}$ coexistence is established.
Finally, the $\phi$ range of the $\m{CsCl}$ phase becomes smaller upon increasing $\kappa a$ 
as can be inferred from Figs.~\Fig{Tpphi3}(f) and \Fig{Tpphi10}(f).
   
In order to make predictions on the conditions to synthesize certain crystal structures, it 
is expected within the present formalism that colloids with strongly screened Coulomb 
interaction, i.e., large values of $\kappa a$, are preferable to prepare $\m{CuAu}$ structures, 
whereas $\m{CsCl}$, $\m{Cu_3Au}$, and $\m{NbP}$ crystals are expected to be found most easily 
in systems of weakly screened Coulomb interaction.
Given a certain crystal structure has been prepared, the above reasoning leads to the 
following conclusions: $\m{CuAu}$ structures become more whereas $\m{CsCl}$, $\m{Cu_3Au}$, and
$\m{NbP}$ structures become less stable against temperature variations upon increasing 
$\kappa a$.

%-------------------------------------------------------------------------------

\section{\label{sec:summary}Summary}
  
In this work a simple, yet semi-quantitative mean-field theory for binary colloidal mixtures 
of spherical particles of equal radii and opposite charges has been described and applied to
calculate the global phase diagrams in terms of temperature, pressure and composition for 
various screening strengths (see Figs.~\Fig{Tpphi3}--\Fig{Tpphi10}).
The formalism interpolates between the hard-sphere limit at high temperatures and a 
Madelung-like description at low temperatures.
The reliability of the method has been checked by comparison with computer simulation results 
\cite{Hynninen2003,Hynninen2006a} for one single composition and screening strength as well as with 
Madelung sum calculations \cite{Leunissen2005,Maskaly2006,vdBerg2009} in the low-temperature limit. 
Limitations of the present formalism can be understood as a result of the absence of 
fluctuations within mean-field theories and properties of the mean-spherical approximation (MSA). 
The influence of the screening strength on the stability of crystalline phases has been 
discussed, which has implications, e.g., on the preparation procedure and the choice of 
experimental conditions.

An extension of the presented theory to multi-component, size- and charge-polydisperse 
colloidal suspensions along the lines described in Sec.~\ref{sec:formalism} is straightforward
as the free volume equation of state used for the solid phases does not explicitly depend on 
the colloid radii and more general formulations of the MSA are available 
\cite{Blum1992,Ginoza1998}.
It is therefore a matter of choosing an appropriate set of candidate solid structures 
(see Tab.~\ref{tab:candidates}), which could be motivated by findings from experiments and 
computer simulations of special cases.

%-------------------------------------------------------------------------------

\begin{acknowledgments}
   MB thanks A.\ Macio{\l}ek for helpful comments. 
   MD acknowledges financial support from an NWO-vici grant.
   This work is part of the research program of the ``Stichting voor
   Fundamenteel Onderzoek der Materie (FOM)'', which is financially supported by
   the ``Nederlandse Organisatie voor Wetenschappelijk Onderzoek (NWO)''.
\end{acknowledgments}

%-------------------------------------------------------------------------------

%-------------------------------------------------------------------------------

\end{document}